\documentclass[a4paper]{spie}  
\usepackage[]{graphicx}

\title{MICADO: the E-ELT Adaptive Optics Imaging Camera}

\author{Richard Davies\supit{a}, 
N.~Ageorges\supit{a}, 
L.~Barl\supit{a},
L.~R.~Bedin\supit{j},
R.~Bender\supit{a,c},
P.~Bernardi\supit{h},
F.~Chapron\supit{h},
Y.~Clenet\supit{h},
A.~Deep\supit{d},
E.~Deul\supit{d},
M.~Drost\supit{f},
F.~Eisenhauer\supit{a},
R.~Falomo\supit{g},
G.~Fiorentino\supit{e},
N.~M.~F\"orster~Schreiber\supit{a},
E.~Gendron\supit{h},
R.~Genzel\supit{a},
D.~Gratadour\supit{h},
L.~Greggio\supit{g},
F.~Grupp\supit{c},
E.~Held\supit{g},
T.~Herbst\supit{b},
H.-J.~Hess\supit{c},
Z.~Hubert\supit{h},
K.~Jahnke\supit{b},
K.~Kuijken\supit{d},
D.~Lutz\supit{a},
D.~Magrin\supit{g},
B.~Muschielok\supit{c},
R.~Navarro\supit{f},
E.~Noyola\supit{a,c},
T.~Paumard\supit{h},
G.~Piotto\supit{g},
R.~Ragazzoni\supit{g},\\
A.~Renzini\supit{g},
G.~Rousset\supit{h},
H.-W.~Rix\supit{b},
R.~Saglia\supit{a},
L.~Tacconi\supit{a},
M.~Thiel\supit{a},
E.~Tolstoy\supit{e},
S.~Trippe\supit{i},
N.~Tromp\supit{f},
E.~A.~Valentijn\supit{e},
G.~Verdoes~Kleijn\supit{e},
and
M.~Wegner\supit{c}
\skiplinehalf
\supit{a}
Max Planck Institute for extraterrestrial Physics,
 Postfach 1312, 85741 Garching, Germany\\
\supit{b}
Max Planck Institute for Astronomy,
K\"onigstuhl 17, 69117 Heidelberg, Germany\\
\supit{c}
Munich University Observatory,
Scheinerstrasse 1, 81679 M\"unchen, Germany\\
\supit{d}
Leiden Observatory, 
Leiden University, 2300RA Leiden, Netherlands\\
\supit{e}
Kapteyn Astronomical Institute, 
University of Groningen, 9700AV Groningen, Netherlands\\
\supit{f}
NOVA-ASTRON, 
P.O.Box 2, Dwingeloo, Netherlands\\
\supit{g}
INAF -- Astronomical Observatory of Padova,
Vicolo dell'Osservatorio 5, 35122 Padova, Italy\\
\supit{h}
LESIA, Observatoire de Paris, 
5 place Jules Janssen, 91195 Meudon cedex, France\\
\supit{i}
IRAM, 
300 rue de la Piscine, 38406 Saint Martin d'H\`eres, France\\
\supit{j}
Space Telescope Science Institute, 
3800 San Martin Drive, Baltimore, MD 21218, USA
}

\authorinfo{Further author information: (Send correspondence to
  R.D.)\\E-mail: davies@mpe.mpg.de}

\newcommand{\arcsec}{\hbox{$^{\prime\prime}$}}
\newcommand{\arcmin}{\hbox{$^{\prime}$}}
\newcommand{\micron}{\,\hbox{$\mu$m}}

\begin{document} 
  \maketitle 

\begin{abstract}

MICADO is the adaptive optics imaging camera for the E-ELT. It has
been designed and optimised to be mounted to the LGS-MCAO system
MAORY, and will provide diffraction limited imaging over a wide
($\sim$1\,arcmin) field of view. For initial operations, it
can also be used with its own simpler AO module that provides
on-axis diffraction limited performance using natural guide
stars. We discuss the instrument's key capabilities and expected
performance, and show how the science drivers have
shaped its design. We outline the technical concept, from the
opto-mechanical design to operations and data processing.
We describe the AO module, summarise the instrument performance, and
indicate some possible future developments.

\end{abstract}




\section{MICADO Overview}

MICADO is the {\em Multi-AO Imaging Camera for Deep Observations},
designed to work with adaptive optics (AO) on the E-ELT. 
It has been optimised for the multi-conjugate adaptive optics
(MCAO) module MAORY\cite{dio10,fop10}; 
but it is also able to work with other adaptive optics
systems, and includes a separate module to provide a single conjugate
adaptive optics (SCAO) capability\cite{cle10} 
using natural guide stars during early
operations (see Section~\ref{sec:ao}).
As this simple AO mode sets low requirements on the telescope
and facilities (e.g. no lasers are required), it is an optimum
choice for demonstrating the scientific capabilities of the E-ELT at
the earliest opportunity.
The optical relay and support structure for SCAO provide the same
opto-mechanical interface as MAORY, and in principle enable MICADO to
be used with other AO systems such as ATLAS\cite{fus10}.
This phased approach means that MICADO will be able to make use of
increasingly sophisticated AO systems as they become available.

MICADO is compact and is supported underneath the AO systems
so that it rotates in a gravity invariant orientation. It is able to
image, through a large number of selected wide and narrow-band near
infrared filters, a large 53\arcsec\ field of view at the diffraction
limit of the E-ELT. MICADO has two arms. The primary arm is a high
throughput imaging camera with a single 3\,mas pixel scale. This arm is
designed with fixed mirrors for superior stability, thus optimizing
astrometric precision. In addition, MICADO will have an auxiliary arm
to provide an increased degree of flexibility. In the current design,
this arm provides (i) a finer 1.5\,mas pixel scale over a smaller field,
and (ii) a 4\,mas pixel scale for a simple, medium resolution, long-slit
spectroscopic capability. In principle the auxiliary arm also
opens the door to other options, including a `dual imager' based
on a Fabry-Perot etalon to image emission line and continuum
wavelengths simultaneously, coronography (perhaps implemented in a comparable
way to that in NACO\cite{boc04}), or a high time resolution detector.

\begin{figure}
\begin{center}
\includegraphics[height=6cm]{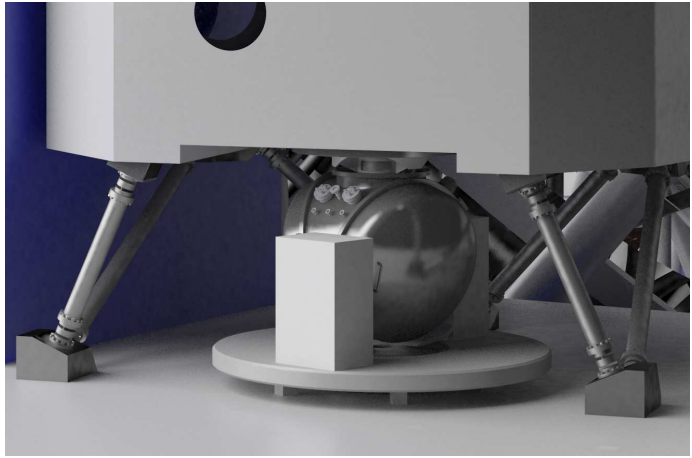}
\end{center}
\caption{Illustration of MICADO mounted under the
  MCAO system MAORY on the E-ELT Nasmyth platform.
\label{fig:photo}
}
\end{figure}

\section{Key Capabilities and Science Drivers}
\label{sec:drivers}

MICADO will excel at several key
capabilities that exemplify the unique features of the E-ELT.
These are at the
root of the science cases, which span key elements of modern
astrophysics,  and have driven the design of the camera.
The science cases are developed in detail elsewhere\cite{ren09} and
here we focus on how MICADO's characteristics enable it to address
them.

\subsection{Sensitivity and Resolution}

MICADO is optimised for imaging at the diffraction limit, and will
fully sample the 6--10\,mas FWHM in the {\it J--K} bands. With a throughput
exceeding 60\% its sensitivity at 1--2\micron\ will be comparable to, or
surpass, JWST for isolated point sources. MICADO's resolution means that
it will be clearly superior to JWST in crowded regions. In addition,
its field of view of nearly 1\,arcmin yields a significant multiplex
advantage compared to other ground-based cameras on ELTs. 
Together, these characteristics make MICADO a powerful tool for
many science cases. Continuum and emission line mapping of high
redshift galaxies will enable it to address questions
concerning their assembly, and subsequent evolution in terms of
mergers, internal secular instabilities, and bulge growth. The
resolution of better than 100\,pc at $z\sim2$, equivalent to 1\arcsec\ imaging
of Virgo Cluster galaxies, will resolve the individual star-forming
complexes and clusters, which is the key to understanding the
processes that drive their evolution. Alternatively, one can probe a
galaxy's evolution through colour-magnitude diagrams that trace the
fossil record of its star formation. Spatially resolving the stellar
populations in this way is a crucial ability, since integrated
luminosities are dominated by only the youngest and brightest
population. MICADO will extend the sample volume from the Local Group
out to the Virgo Cluster and push the analysis of the stellar
populations deeper into the centres of these galaxies.

\subsection{Precision Astrometry}

With only fixed mirrors in its primary imaging field, gravity invariant
rotation, and HAWAII-4RG detectors (developed to meet the
stringent requirements of space astrometry missions), 
MICADO is an ideal instrument for astrometry. A robust pipeline will
bring precision astrometry into the mainstream. An analysis of the
statistical and systematic effects\cite{dav10,tri10} shows that
an accuracy of 40\,$\mu$as in a single epoch of observations is achievable;
and after only 3--4\,years it will be possible to measure proper motions
of 10\,$\mu$as\,yr$^{-1}$, equivalent to 5\,km\,s$^{-1}$ at 100\,kpc. 
At this level, many astronomical objects are no longer static but become
dynamic, leading to dramatic new
insights into the three dimensional structure and evolution of many
phenomena.
Proper motions of faint stars within
light-hours of the Galactic Center will measure the
gravitational potential in the relativistic regime very close to the
central black hole, and may also reveal the theoretically predicted
extended mass distribution from stellar black holes that should
dominate the inner region. The internal kinematics and proper motions
of Globular Clusters will yield insights on intermediate mass black
holes as well as the formation and evolution of the Galaxy. Similar
analyses of Dwarf Spheroidals will reveal the amount and distribution
of dark matter in these objects, and hence test models of hierarchical
structure formation.

\subsection{High Throughput Spectroscopy}

Spectroscopy is an obvious and powerful complement to pure imaging,
and is implemented as a simple slit spectrometer with a high
throughput that is ideal for obtaining spectra of compact objects. The
resolution of $R\sim3000$ is sufficient to probe between the near infrared
OH lines. This simple addition will enhance many science 
cases, for example: deriving stellar types and 3D orbits in the
Galactic Center; using velocities of stars in nearby galaxies to probe
central black hole masses and extended mass distributions; measuring
absorption lines in galaxies at $z=2$--3 and emission lines in galaxies
at $z=4$--6 to derive their ages, metallicities, and star forming
histories; and obtaining spectra of the first supernovae at $z=1$--6.

\section{Technical Design Concept}
\label{sec:design}

The MICADO design to achieve the capabilites outlined in
Section~\ref{sec:drivers} is simple, compact, and
robust. As such, it minimizes risks on cost and schedule.
In the following, we outline the key features of its design.

\subsection{Optics}
\label{sec:optics}

\begin{figure}
\begin{center}
\includegraphics[height=6cm]{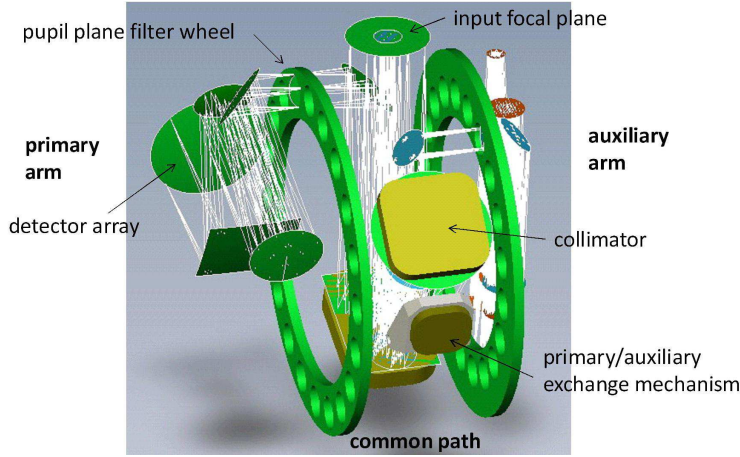}
\end{center}
\caption{Overview of the MICADO optics.
The major components are labelled.
\label{fig:optics}
}
\end{figure}

The optics are discussed in more detail elsewhere\cite{mag10} and so
only a brief outline of the main characteristics is given here.
The design was reached after a scientific and technical trade-off.
The main requirements from this are that: 
the pixel scale in the primary arm should be fixed to maximise
stability; 
the scale should be 3\,mas to cover a large field of view while being
Nyquist sampled in {\it J}-band;
there should be space for a large number of filters;
the degree of distortion is less important than its stability (since it must be corrected anyway);
the instrument should cope with a strongly curved input wavefront, as
well as a flat input wavefront (for a reduced field of view, limited
by anisoplanatism, during SCAO operations).

\begin{table}
\begin{center}
\caption{Characteristics of the optical designs for the 2 arms}
\label{tab:optics}
\begin{tabular}{|p{2.7cm}|p{6.5cm}|p{6.5cm}|}
\hline

&
\multicolumn{1}{c|}{Primary Arm} & 
\multicolumn{1}{c|}{Auxiliary Arm}
\\
\hline

Throughput &
61--70\% for Y--K bands &
60--69\% for imaging in Y--K bands;\newline 
18--28\% for spectroscopy in Iz--K bands
\\
\hline

FoV \& pixel scale &
53\arcsec$\times$53\arcsec at 3\,mas/pixel &
6.4\arcsec$\times$6.4\arcsec\ at 1.5\,mas/pixel (imaging);\newline
17.1\arcsec$\times$17.1\arcsec\ at 4\,mas/pixel (spectroscopy)
\\
\hline

Filters &
Single wheel with 20 positions for filters &
Single wheel with 20 positions for filters and grisms
\\
\hline

Image quality &
Nominal Strehl ratio at 0.8\micron\ is $>$83\% across the whole field &
$>$89\% for 1.5\,mas imaging;\newline $>$75\% in slit for 4\,mas scale.
\\
\hline

Distortion &
1.2\% across the whole field &
-0.39\% at 4\,mas; 0.03\% at 1.5\,mas
\\
\hline

Optical Ghosts &
\multicolumn{2}{p{13cm}|}{The reflective optics do no create ghosts; the ADC,
  entrance window, and filters are tilted so that ghosts on the
  detector are minimised}
\\ 
\hline

Largest Fold\newline Mirror &
\multicolumn{2}{p{13cm}|}{260$\times$380\,mm (in common path)}
\\
\hline

Largest Working\newline Mirror &
256$\times$276\,mm &
200\,mm diameter (in common path)
\\
\hline

Tolerances &
\multicolumn{2}{p{13cm}|}{0.05\,mm and 0.01$^\circ$ (i.e. within
manufacturable limits) for 70\% strehl at 1\micron}
\\
\hline

Intermediate Pupil &
Shape varies with field position so
cold stop is undersized at 99.1\,mm. Maximum vignetting is 1.0\% in corner of
field &
Cold stop undersized at 85.6\,mm.\newline Maximum vignetting is 0.4\%
\\
\hline

Focal Plane &
264$\times$268\,mm; tilted by 4.1$^\circ$ and
convex with a 1500\,mm curvature radius (due to input
wavefront curavature) &
61.4\,mm across;\newline unaffected by input curvature
\\
\hline

SCAO image\newline quality &
Strehl ratio is $>$84\% at 0.8\micron\ over full 27\arcsec\ SCAO field &
$>$78\%
\\

\hline
\end{tabular}
\end{center}
\vspace{-5mm}
\end{table}

As shown in Fig.~\ref{fig:optics} the MICADO optics comprises 3
sub-systems: the common path, primary arm, and auxiliary arm.
The first component in the common path is a tunable atmospheric
dispersion corrector (ADC; not shown in the figure).
It consists of 2 pairs of ZnSe/ZnS prisms that can
be rotated to provide a dispersion correction that is optimised to
the observational band. 
It is located as far in front of the input focal
plane as possible (i.e. 500mm).
This is to minimize its impact on the optical quality by enabling
thinner prisms with smaller wedge angles to be used. 
In its current location (warm and right against the interface to
MAORY), the performance is just acceptable. 
However, this location is not optimal and during Phase B other
options, such as locating it at an appropriate pupil plane within
the AO system, will be addressed. 

Both arms use an off-axis parabola for collimation. 
However, the sizes and locations of the parabolae are different, and
so the primary arm has a fixed mirror, while the auxiliary 
arm requires an alternative mirror to be rotated into position. 
In both cases, to keep the optical system compact, the light is
reflected in both directions from a large fold mirror. 
Separate fixed fold mirrors then direct the light out of the
common path to opposite sides of the instrument. 
The collimator creates a pupil image just after this fold mirror,
where a large filter wheel is located. 
The maximum circular diameter of the pupil is 100.5\,mm; 
however, in order to block unwanted thermal background (the pupil has a shape
that depends slightly on field position), the coldstop is
undersized at 99\,mm diameter. 
The primary arm has only 3 additional working mirrors (based on a
TMA), although extra fold mirrors are required to keep the volume
occupied small.
The auxiliary arm is similar, but includes mechanisms for
changing the pixel scale.
Table~\ref{tab:optics} summarizes the key characteristics of the two arms.

The baseline detector for MICADO is the HAWAII-4RG.
These have a number of important advantages: 
(i) they are large format, so that relatively few detectors need to be
characterised and mosaiced,  
(ii) they have been designed for the stringent requirements of space
astrometry missions and so are ideal for MICADO's astrometry
applications,  
(iii) the readout speed can be adjusted (even on multiple
sub-regions), greatly reducing the impact of saturation due to bright
targets. 
The cross-talk between pixels is relatively low and electronic ghosts
can largely be suppressed in the same way as is done for XShooter.
There will be relatively small gaps (each a few~mm compared to the
25\,cm width of the focal plane) between individual detectors since
they are not directly buttable. 
This can be considered advantageous since it provides a
quasi-coronagraph, allowing one to position a bright star out of the
field of view even in dithered exposures.  
To optimise stability of the focal plane array, all detectors will be
mounted on a single baseplate. 
A direct result of the strongly curved input focal plane from the MCAO system
(which cannot be corrected in a satisfactory way) is that the MICADO focal
plane is tilted by 4.1$^\circ$ and curved, with a 1500\,mm
radius of curvature.
Because the 60\,mm wide detectors are flat, there will be a small
defocus across each detector due to the $\pm$0.32\,mm 
focal plane mismatch. 
This has only a minor impact on the spot diagrams, and the
Strehl ratio at 0.8\micron\ remains above 88\% across the whole field
(i.e. significantly larger than that used for
tolerancing).

\subsection{Mechanics}
\label{sec:mech}

\begin{figure}
\begin{center}
\includegraphics[height=6cm]{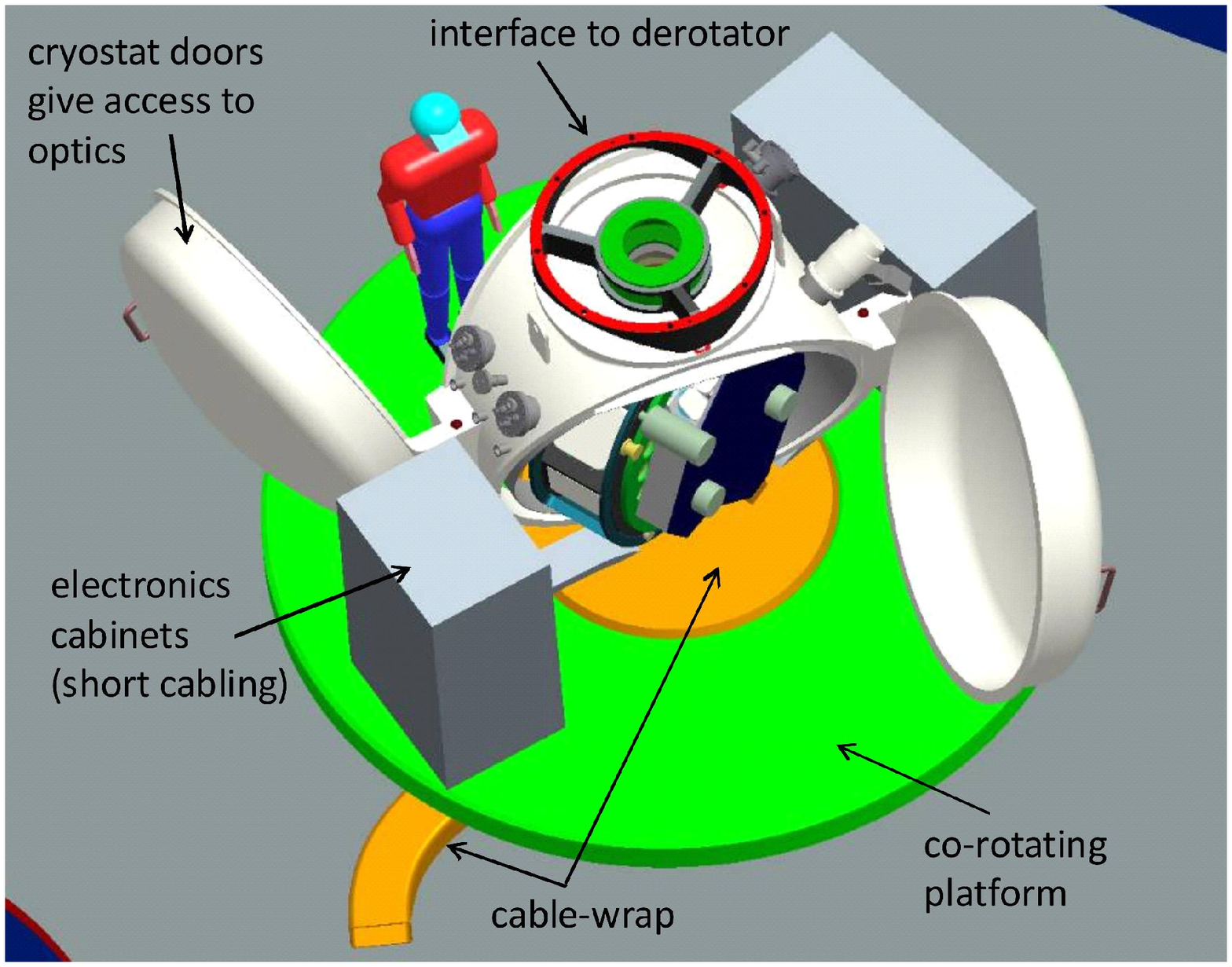}
\includegraphics[height=6cm]{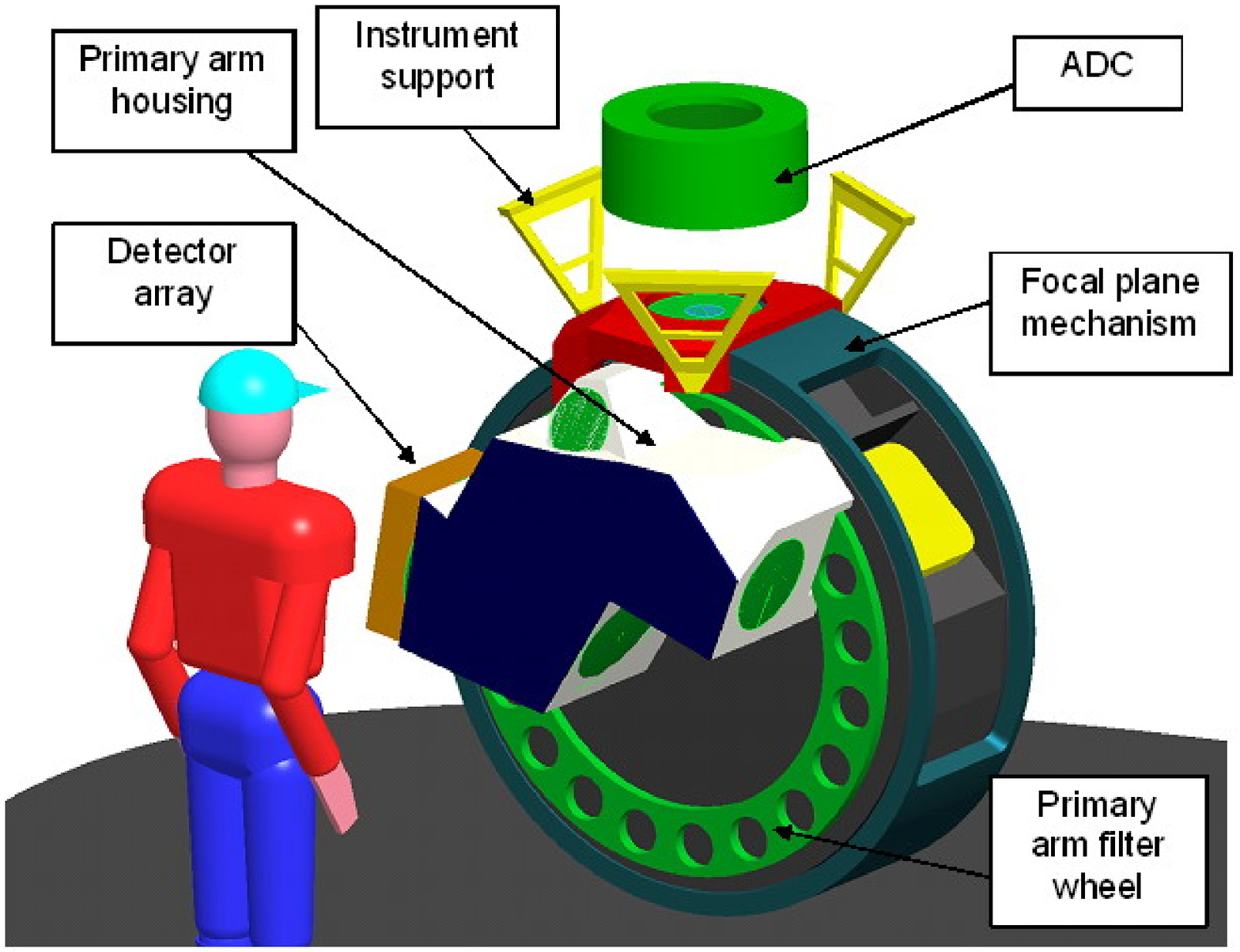}
\end{center}
\caption{Left: The main components of MICADO, with the cryostat access doors
  open.
Right: overview of the MICADO mechanical structure, with the main
  structures labelled.
\label{fig:mech}
}
\end{figure}

The mechanical design, and folding of the optical path, has
largely been driven by the limited space under MAORY. 
To keep torques small and to maintain optical alignment during
cool-down, the centre of gravity is close to the optical axis, which
itself is close to the centre of shrinkage. 
In order to minimize cable lengths and to limit the mass mounted on
the derotator, the electronics racks stand on a co-rotating
platform supported on the Nasmyth floor.
This platform also houses the cable-wrap for external supplies. 
Service and maintenance are also key aspects,
leading to a design in which the core instrument and optics structure
are rotated by 25$^\circ$ with respect to the cryostat. 
This provides better access through the cryostat doors to the detector
arrays, the arm selection and focal plane
mechanisms, the filter wheels, and the core optics.

MICADO is housed in a stainless steel cryostat (Fig.~\ref{fig:mech}, left),
which has a fixed tapered part with sufficient space for all the
through-ports and pumps. 
On either side are 2 large doors which provide access to all key
components while MICADO is mounted to the AO system. 
The 3 electronics cabinets (2 of which are back-to-back) are
positioned on the co-rotating 
platform in such a way that they do not interfere with the doors.
The entrance window of the cryostat is located 300\,mm above the focal
plane and 200\,mm below the mounting interface. 
The warm ADC is currently located in this volume for the reasons
outlined in Section~\ref{sec:optics}. 
To minimise flexure (which is critical for astrometry), the instrument
has been designed for gravity invariant rotation. 
Because of this, and since the cool-down times are limited by
thermal contact of the filters, the cryostat has not been
light-weighted.
The total mass of the cryostat and instrument supported by the
derotator is 3000\,kg.
An additional 2800\,kg are supported on the Nasmyth floor, and 500\,kg
in a calibration unit located in the AO system.

Supported inside the cryostat behind the radiation shield is the cold optics
instrument (Fig.~\ref{fig:mech}, right).
It comprises 3 main structures:
the primary arm, auxiliary arm, and core sub-assembly. 
The general design approach for each of these housings is to assemble
them from plate material to keep part complexity and accuracy low, and
ensure a rigid boxed structure. 
Stray light can be reduced by proper shielding and baffling, and also
by using a wave-like finish to the surface of walls, and applying a
low-reflectivity black coating. 
The filter wheels for each of the arms are mounted on the side
panels of the core structure, and supported at their perimeter. 
The primary and auxiliary arm housings are mounted to the core with a
three-point interface in such a way that they connect to the round
side panels with direct support underneath from the core wall panels.
The instrument core is supported by the cryostat via 3 V-rods and a
transfer structure.
The rods are co-axially aligned to maintain alignment during
cool-down, and are attached to the transfer structure. 
This has been designed to accommodate the rotating focal plane
mechanism, and acts as a bridge to the stiff support structure of the
core subassembly.
MICADO has relatively few cold mechanisms.
In order to achieve high repeatability, these are all rotational and
use spring loaded bearings and V-grooves to locate and lock each
position:\vspace{-2mm}
\begin{description}
\item{\em Focal plane selection}: the input focal plane is large, and
6 positions are required (field stop for each arm, 2 long slits, a
closed position, and a point source mask for initial check of
internal focus). 
It is a large structure that is driven around the outside of the core
sub-assembly. \vspace{-1.5mm}
\item{\em Primary/Auxiliary Arm selection}: the core contains only
one mechanism, which rotates in an alternative parabolic
collimator mirror if the auxiliary arm is selected. 
Because this is recognised as being delicate, accessibility is
an important design driver, and is facilitated by a large opening in
the focal plane wheel.\vspace{-1.5mm}
\item{\em Filter wheels}: The pupil sizes are 100\,mm and 86\,mm
diameter for the primary and auxiliary arms. 
In order to provide space for 20 filter slots, these wheels are large,
and hence supported and driven at their perimeter. 
Since the filters will dictate the cool-down time, care has been given
to maximising the thermal contact.\vspace{-1.5mm}
\item{\em Scale changing mechanism}: the auxiliary arm has two
mechanisms to move mirrors and enable a pixel scale change. 
The design is similar to the primary/auxiliary selection mechanism.
\end{description}

\subsection{Cryogenics}
\label{sec:cryo}

\begin{figure}
\begin{center}
\includegraphics[height=5cm]{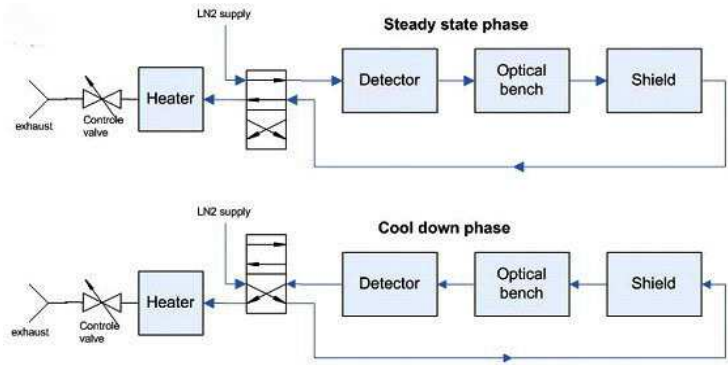}
\end{center}
\caption{Overview of the MICADO reversible cryogenic scheme
\label{fig:cryo}
}
\end{figure}

In order to avoid the use of cryo-coolers, the vibrations from which
would have a strong adverse effect on the AO performance, MICADO will
be cooled by continuous flow liquid nitrogen (LN2) during
cool-down/warm-up cycles as well as steady state phases. 
For cooldown, an estimated 1000\,L will be needed; and to maintain
steady-state the required flow rate, including some contingency, is
expected to be 72\,L/day. 

Continuous flow is preferred over a LN2 bath since it gives more
freedom in the location of the detectors, it keeps the cryostat
smaller and its mass lower, and it is possible to combine the precooling
and steady state systems. 
Cooling pads are located at strategic points in the cryostat, and
connected so that during a cool-down cycle the heat shield is cooled
first, followed by the optical bench and finally the detectors. 
During steady state the sequence is reversed so that the detectors
have the lowest possible temperature. 
This series concept is shown in Fig.~\ref{fig:cryo}, which
demonstrates how the same circuit can be used during both phases. 

\subsection{Electronics}
\label{sec:elec}

The electronics cabinets are physically located very close to the
MICADO cryostat, on a platform that co-corates with it.
This means that the cables can be kept short.
At the current time, no ELT electronics standards have been defined.
The preferences for MICADO include the architecture being based on
SIMATIC Programmable Logic Controller (PLC), Realtime Ethernet or
other Realtime architectures.
A fail-safe version of the PLC SMATIC 57 is also an option for
cryogenic housekeeping (cryogenics control).
And Realtime LabView and PXI controller from National Instruments could
also be used in the implementation of control electronics.

\subsection{Instrument Control}
\label{sec:sw}

\begin{figure}
\begin{center}
\includegraphics[height=8cm]{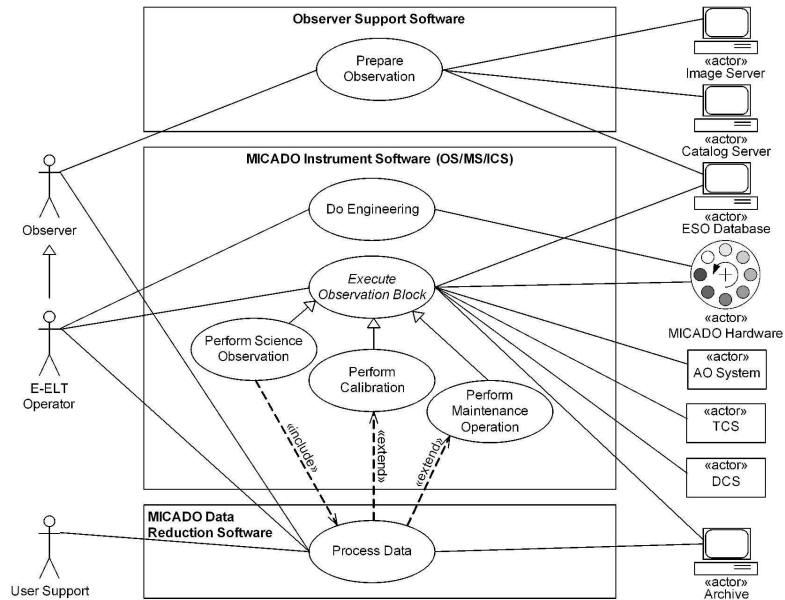}
\end{center}
\caption{Overview of top-level use cases and actors for the MICADO instrument control software
\label{fig:inssw}
}
\end{figure}

The user requirements for the instrument software have been developed
for observation preparation, science operations (including on-sky
calibration), and maintenance operations. 
The main functionality has been analysed via specific use cases
related to the various observing scenarios and modes. 
An overview of the complete scheme is given in Fig.~\ref{fig:inssw}. 
Specific use cases include:\vspace{-3mm}
\begin{description}
\item{\em Science Observations}: 
imaging, spectroscopy, on-sky calibrations
(e.g. standard stars, twilight flats)\vspace{-1.5mm}
\item{\em Calibrations}: 
dark frames, internal flatfield, linearity,
wavelength \& distortion calibrations, ghost
assessment\vspace{-1.5mm}
\item{\em Maintenance Operations}: telescope focus.\vspace{-1.5mm}
\end{description}

It is likely that, apart 
from the actual instrument control software itself, MICADO will share a 
common software layer with other E-ELT subsystems. 
However, at the current time, no E-ELT software standards have been
defined.
Given the timeline of 
the project and the need for a stable development infrastructure, open 
source solutions are preferred over proprietary commercial ones in
this context.
For the same reason, Linux is favoured as the operating system.

\subsection{Data Processing}

The user requirements for the data reduction software have been
developed from observational scenarios for imaging and
spectroscopy. 
These are standard techniques, and lead to no surprises. 
The performance required for photometry and astrometry have led to
additional requirements. 
These include the use of a special internal calibration mask to
measure instrument distortions, and additional steps in the data
processing for the related science projects.
The Astro-WISE system is well suited for reduction of MICADO
data. Astro-WISE is an integrated system where users cannot only
perform data reduction but also data archiving, post-reduction analysis
and publishing of the raw, intermediate and final data products.
A salient feature for the data reduction relevant for MICADO is that
it performs `global' astrometry and photometry. 
Astrometric and photometric corrections and calibrations by combining
the information from overlapping observations improving on
calibrations based on individual pointings. 
Data reduction can take place in fully automated fashion or in a more
manual fine-tuning manner. 
The data rates estimated for MICADO are up to about 6\,Terabytes per
night (if all individual exposures are kept and processed for optimal
astrometric accuracy); 
although significantly less if either short exposures can be directly
co-added in the detector control system (DCS), or longer exposures are
required (e.g. when using narrow band filters).

\section{Adaptive Optics}
\label{sec:ao}

The design of MICADO has been optimised for the multi-conjugate
adaptive optics module MAORY\cite{dio10,fop10}, which uses multiple
lasers and natural guide stars to provide
diffraction limited performance over a wide field with high sky coverage.
However, the Phase A study included a simpler AO system that can be
used during initial operations, in order to mitigate the risk
associated with such a complex AO system, and to enable MICADO
to produce diffraction limited images at the earliest opportunity.
This AO system will be single-conjugate and use a single natural guide
star as the wavefront reference:
there will be sufficient science targets near suitable
guide stars for 2--3 years of operation.
Including SCAO in the MICADO design is necessary
because, at the current time, the E-ELT baseline does not include a
wavefront sensing capability for scientific instruments (although any
WFS can make use of the E-ELT's deformable and tip-tilt mirrors).
And because
MICADO cannot interface to the Nasmyth port, a major part of the SCAO
system is an optical relay and support structure that provides the
same mechanical and operational interface as to MAORY.
Since it can in principle be re-used with other AO
systems such as ATLAS\cite{fus10}, 
this means that MICADO will be able to make use of
increasingly sophisticated AO systems as they become available.

\begin{figure}
\begin{center}
\includegraphics[height=6cm]{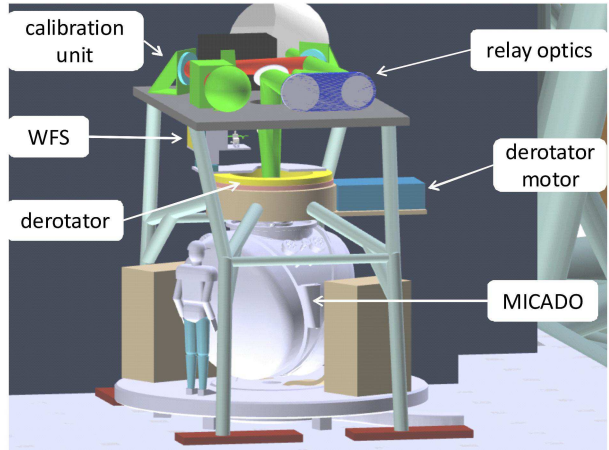}
\end{center}
\caption{MICADO SCAO module `SAMI'.
\label{fig:scao}
}
\end{figure}

The top level requirements for the SCAO module are relatively simple:
the optical, mechanical, and communication interfaces
should be the same as those to/from MAORY;
the WFS bandpass should be 0.45--0.8\micron, to maximise sensitivity without compromising the scientific wavelength range;
the WFS should be able to guide on stars anywhere within a
45\arcsec\ diameter patrol field;
and the transmitted scientific field of view need only be
27\arcsec$\times$27\arcsec\ (commensurate with the isoplanatic patch
size).
This last requirement means that even though there is no field
curvature from the telescope (in contrast to the strongly curved field
from MAORY for which MICADO is designed), the image
quality over the SCAO field is unaffected.
It also means that initially, only the
central detectors need be mounted; the remainder can be
integrated later.
SAMI, the SCAO module for MICADO shown in Fig.~\ref{fig:scao}, is
described in detail elsewhere\cite{cle10}. 
It comprises 4 sub-systems:\vspace{-2mm}
\begin{description}
\item{\em an optical relay} made of a 3-mirror Offner relay, a folding
  mirror directing the light downward to MICADO and the WFS, and a
  dichroic plate splitting the light between MICADO and the
  WFS.\vspace{-1.5mm}
\item{\em a field derotator} to compensate for the telescope movements
  while tracking, for both MICADO and the WFS.\vspace{-1.5mm}
\item{\em a support structure} for the
  optical bench of the relay optics, the WFS, the derotator and
  MICADO. \vspace{-1.5mm}
\item{\em the WFS}, including all opto-mechanics after the
  dichroic. It comprises a pupil steering mirror, and, mounted on XY
  stages, a field stop, a K-mirror for
  pupil derotation, a lens triplet and the WFS camera itself.
The initial study suggests that an ADC
  in the WFS should not be necessary.\vspace{-1.5mm}
\end{description}

The performance of the MICADO SCAO module has been estimated using
analytical formulae (e.g. for the anisoplanatism error), information
from ESO (e.g. for the fitting error) and two home-made simulation tools. 
It takes into account the control laws that will be implemented: a
classical integrator, with a modal control, together with a Kalman
filter for windshake compensation; 
and also that smart algorithms, such as weighting or pixel selection,
will be used for the centre of gravity computation.
The results of these computations are summarized in 
Tables~\ref{tab:scao} and~\ref{tab:aniso}.
Multiplying values from both tables together will yield an estimation
of the strehl ratio expected for a guide star of a given magnitude at
a given distance off-axis.

\begin{table}[h]
\begin{center}
\caption{SCAO performance as a function of guide star magnitude}
\label{tab:scao}
\begin{tabular}{|l|c|c|c|c|c|c|}
\hline
On-axis reference source V-band magnitude & 
12 & 13 & 14 & 15 & 16 \\
\hline
Total wavefront error (nm rms) & 
183 & 205 & 245 & 328 & 514 \\
\hline
Strehl at 2.2\micron & 
76\% & 71\% & 61 \% & 41\% & 12\% \\
\hline
\end{tabular}
\end{center}
\vspace{-5mm}
\end{table}

\begin{table}[h]
\begin{center}
\caption{Anisoplanatic effect on strehl ratio for SCAO}
\label{tab:aniso}
\begin{tabular}{|l|c|c|c|c|c|c|}
\hline
Distance from reference source (arcsec) & 
5 & 15 & 25 & 35 & 45 & 55 \\
\hline
Anisoplanatism error for L0=25m (nm rms) & 
101 & 253 & 354 & 433 & 518 & 554 \\
\hline
Corresponding Strehl ratio scaling at 2.2\micron & 
92\% & 59\% & 36 \% & 22\% & 11\% & 8\% \\
\hline
\end{tabular}
\end{center}
\vspace{-5mm}
\end{table}

\section{Operation and Calibration}
\label{sec:ops}

The basic operational scenario for MICADO is very similar to other
imaging cameras and spectrometers.
For imaging, the sky background will be derived either by combining
dithered exposures or, when necessary, by offsetting to sky. 
For spectroscopy, the source will be nodded back and forth along the
slit. 
Typical exposure times will be a few seconds (broad band filters) up
to tens of seconds (narrow band filters). 
For the shortest exposure times, several exposures will be made at the
same pointing before dithering.
The main issue is the size of the dithers, which must be optimised for
science while minimizing the AO and telescope overheads. 
Table~\ref{tab:dither} summarises the definition of dithers and
offsets for MICADO.

\begin{table}[h]
\begin{center}
\caption{Definition of dithers and offsets to reduce AO and telescope overheads}
\label{tab:dither}
\begin{tabular}{|l|p{14cm}|}
\hline
Small dither & 
Offset of up to $\pm$0.3\arcsec\ (goal$\pm$0.5\arcsec)
from initial pointing in each of X- and Y- directions, with an
accuracy of $<$2\,mas. 
AO loops remain closed during operation. 
Cadence: 10--30\,sec.\\
\hline
Large dither & 
Offset of up to $\pm$10\arcsec\ from the initial pointing.
AO loops open during the offset, but reclose at the new position.
The telescope is involved.
Cadence: a few minutes.\\
\hline
Sky Offset &
Offset of up to 15\arcmin\ (when background cannot be recovered by
dithering).
AO loops do not need to close in the new position.
Cadence: 10--30\,minutes (depending on overhead).\\
\hline
Sky Return &
Offset back, after a `Sky Offset'.
AO loops should reclose.\\
\hline
\end{tabular}
\end{center}
\vspace{-5mm}
\end{table}

Most of the calibrations can be performed internally during the day
while the dome lights are on: flatfields, wavelength calibration,
darks. 
Additional twilight flats will be required in order to correct
illumination gradients in the internal flats.
The only non-standard calibration required is that to correct
instrument distortions in the AO system and MICADO. 
This will also be possible during the day with the dome lights on, and
will be achieved by inserting a special calibration mask into the
focal plane in front of the AO system. 
The only standard nighttime calibration is to observe standard stars
for flux calibration.

\subsection{Astrometric Calibration}

\begin{figure}
\begin{center}
\includegraphics[height=6cm]{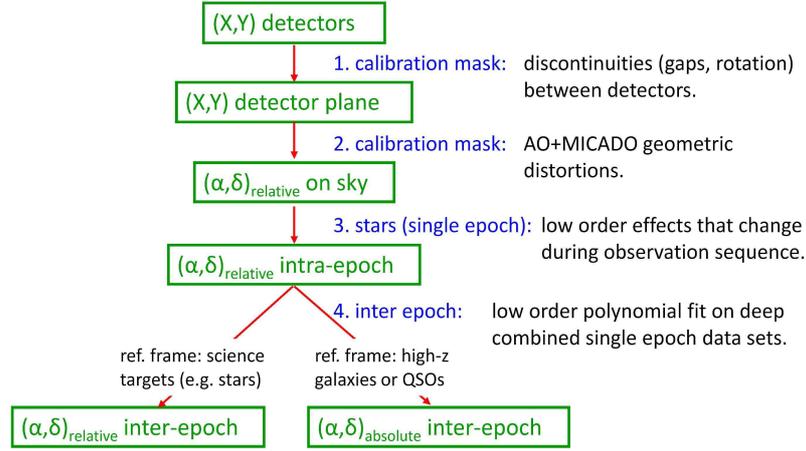}
\end{center}
\caption{Astrometric calibration for MICADO is achieved in several
  stages.
\label{fig:astrocal}
}
\end{figure}

In the Galactic Centre, it is possible to achieve a relative astrometric
precision of 200--300\,$\mu$as\cite{fri10} in the {\it H}-band on an 8-m
class telescope.
This corresponds to about 0.5\% of the FWHM of the PSF.
If this performance is projected forward to the E-ELT, one can hope
to reach a precision of about 40\,$\mu$as.
In Table~\ref{tab:astrocal} we summarise the conclusions from a
study\cite{tri10} of the ten error sources that need to be controlled in
order to achieve this.
Measuring these effects clearly requires careful calibration, and 
the scheme outlined in Fig.~\ref{fig:astrocal} shows how this can be
done.
An internal calibration mask is used to measure discontinuities
between the detectors and instrumental distortions in order to map the
detector plane onto the sky.
This provides a set of relative (or artifical) coordinates to which subsequent
exposures can be matched by applying a low order transformation based on
point sources in the field.
Exposures within a single epoch can then be combined.
Deep integrations obtained in this way at different epochs are mapped
to each other via another low order transformation, using either faint
compact galaxies (which have negligible proper motions) or by using
the ensemble of stars. 
In the latter case, their relative proper motions (a high order
effect) are preserved during a low order transformation.

\begin{table}[h]
\begin{center}
\caption{Sources of astrometric uncertainty}
\label{tab:astrocal}
\begin{tabular}{|p{3.5cm}|p{12.5cm}|}
\hline
Absolute plate scale & 
pre-imaging (seeing limited) can give better than 10$^{-4}$ accuracy
(i.e. 5\,mas over 50\arcsec\ field).\\
\hline
Sampling \& pixel scale & 
no measurable errors if pixel scale does not exceed
3\,mas\,pix$^{-1}$ (note that a finer scale is beneficial for highly
crowded fields).\\
\hline
Instrumental\newline distortions &
are measured to 0.01\,pixel in many current instruments.
For MICADO, calibration with an internal mask will reduce this error to
$\sim$30\,$\mu$as.\\
\hline
Telescope instabilities &
plate scale, rotation, etc., are low order effects that can be
absorbed into a coordinate transformation.\\
\hline
Achromatic differential refraction & 
this large $\sim$10\,mas effect is linear and so is
removed by a coordinate transformation.\\
\hline
Chromatic differential refraction & 
produces $\sim$1\,mas scale effects that depend on the source
colour. 
A tunable ADC can reduce it to $<$20\,$\mu$as in most cases.\\
\hline
AO instrumental \&\newline atmospheric effects & 
shift the relative
positions of the NGS used by the AO system.
MAORY uses 3 NGS and so this is expected to be a low-order effect.
Since they are slow effects, they can also be removed by tracking the
barycentre of each NGS.\\
\hline
Differential tilt jitter & 
introduces errors of $\sim$100\,$\mu$as into diffraction limited E-ELT
observations. 
It scales as $t^{-1/2}$ and can be integrated down to $\sim$10\,$\mu$as
within about 30\,min.\\
\hline
PSF variations & 
even with MCAO the PSF changes across the field of view.
Measurements on simulated PSFs indicate this should
intrdouce errors of $<$10\,$\mu$as.\\
\hline
Galaxies as astrometric references & 
galaxies are spatially resolved, but making use of their detailed internal
structure enables one to reach $\sim$20\,$\mu$as accuracy with deep
integrations.\\
\hline
\end{tabular}
\end{center}
\vspace{-5mm}
\end{table}

\subsection{PSF Calibration and Photometry}

The precision with which photometry can be performed is determined by the
accuracy to which the PSF is known.
This issue is being addressed in 2 complementary ways.
The MAORY consortium are developing a simple model for the PSF which
enables its shape to be determined with only a few parameters.
In principle, one can map the variation of each
parameter across the field of view.
With relatively few empirical measurements, this might yield a quantitative
estimate of the PSF at any point.
Such a tool would be extremely important for many science cases.
In particular, for studying black hole and host galaxy growth across
cosmic time, an accurate estimate of the PSF is needed in order to
separate the QSO and host galaxy emission. 
In crowded stellar fields, simulations indicate that it is possible to
derive the PSF from the data itself and perform accurate photometry using
currently available tools.
This works over a small field where PSF variations are negligible.
For MCAO, simulations indicate that although spatial variations in the PSF
are small, they will have some impact on photometric accuracy.
Therefore, to cover a larger field, one would either need to stitch together
multiple sub-fields that are analysed separately; or develop the
photometry tools to cope with a spatially variable PSF.

\section{Performance}

\begin{figure}
\begin{center}
\includegraphics[height=6cm]{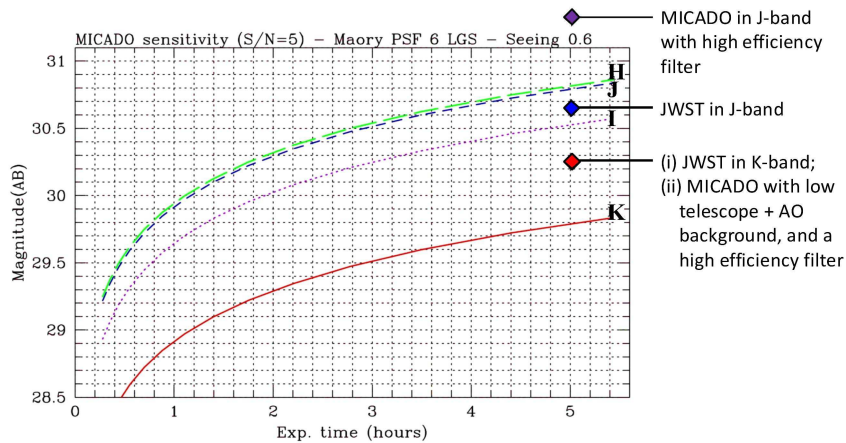}
\end{center}
\caption{MICADO sensitivity (as a function of integration time) for
  broad-band imaging through standard filters. 
A few reference points for 5\,hr integrations
  are shown for comparison.
\label{fig:sens}
}
\end{figure}

The broadband imaging performance for the MICADO primary field is show
in Fig.~\ref{fig:sens}. 
This has been calculated for isolated point sources using
PSFs provided by the MAORY consortium and for standard broadband
filters similar to those in HAWK-I. 
It shows that the 5$\sigma$ sensitivity
will be better than a few nano-Jy (30\,mag AB) for the {\it I--H}
in only 1--2\,hours. The {\it K}-band performance depends strongly on the
thermal background and hence the ambient temperature, but is likely to
be about 1\,mag less.
Advanced filters (see Section~\ref{sec:techno}) will have a very
significant impact on MICADO sensitivity. 
A prototype {\it J}-band filter pair increases the
sensitivity in a given integration time by 0.3\,mag. More
advanced design optimisation techniques could lead to a 0.5\,mag
sensitivity gain in this band, and comparable gains may be expected
for the {\it I}-band and {\it H}-band.

\begin{table}[h]
\begin{center}
\caption{Sensitivity (AB mag) for isolated point sources to 5$\sigma$ in 5\,hours}
\label{tab:sens}
\begin{tabular}{|l|c|c|c|}
\hline
& {\it J$_{AB}$} & {\it H$_{AB}$} & {\it K$_{AB}$} \\
\hline
Imaging & 30.8 & 30.8 & 29.8 \\
Imaging with advanced filters & 31.3 & 31.3 & 30.2 \\
Spectroscopy between the OH lines & 27.2 & 27.2 & 25.7 \\
\hline
\end{tabular}
\end{center}
\vspace{-5mm}
\end{table}

The spectroscopic performance has been calculated for isolated point
sources that are nodded back and forth along a slit that is 8\arcsec\
long and 12\,mas wide. 
Because of the unusually extreme core+halo
shape of the adaptive optics PSF, this width maximises the
signal-to-noise reached for point sources in the {\it J} and {\it H}-bands. 
In the {\it K}-band, additional diffraction losses at the slit reduce the
throughput slightly. The sensitivity calculation takes account of all
effects (including the Strehl ratios predicted by MAORY, the limited
coupling efficiency due to the PSF shape, diffraction losses at the
slit, and the thermal background). The resulting 5$\sigma$
sensitivities are {\it J$_{AB}$=H$_{AB}$}=27.2\,mag between the OH lines in
a 5\,hour integration; 
and similarly {\it K$_{AB}$}=25.7\,mag (which is, as before, less primarily
due to the thermal background).

\section{Technological Developments and Risks}
\label{sec:techno}

MICADO is a simple camera and has been designed to have few
risks. Indeed, the preliminary risk register contains no technical or
programmatic risks above a ‘low’ level. Those that do exist at this
level are common risks associated with all (cryogenic) instruments,
and not specific to MICADO. 
There are several future developments that could be beneficial.
Although none of these is required for the successful functioning
of MICADO, they would each increase the competitiveness of MICADO with
respect to other facilities. The developments include:\vspace{-2mm}

\begin{description}
\item{\em Advanced filters}: 
Substantial gains in sensitivity of ground-based near-infrared
instruments can be attained by sky line suppression or avoidance. 
We have begun a research project with Laser Zentrum
Hannover to develop high throughput broad band filters and
OH suppressing interference filters. The initial work, nearing
completion, is to make a prototype for the {\it J}-band
comprising low-pass and high-pass filters coating opposite sides of a
substrate. 
Together these make a broad-band filter with $>$95\% throughput
(filters with 96--99\% throughput, but suppression over a shorter
baseline, were already maunfactured several years ago\cite{gue08}).
The OH suppression is achieved by transmitting several
narrow bandpasses within this range where the background is
sufficiently low. 
Future development will focus on extension to other bands,
optimisation of the filter profile, process qualification, and coating
homogeneity, stress, and characterisation.

\item{\em Dual Imager}: 
Fabry-Perots are complementary to integral
field spectroscopy, but provide higher quality images (greater
fidelity, and higher resolution over a larger field) of individual
emission lines.
The key to success would be to enable simultaneous imaging of emission
line and continuum wavelengths.
This would avoid problems with variable seeing or AO performance when
subtracting the continuum to obtain the line emissiom map.
Some development is required to achieve a good optical design.

\item{\em High Time Resolution Astronomy}\cite{she08}: 
Scientific applications include the
stochastic behaviour of neutron stars and white dwarf accretion disks,
and pulsar magnetospheres; and time resolved
observations of gamma-ray and X-ray transients and anomalous
repeaters.
Detector technology is available now in the range 0.8--1.2\micron\
using avalanche photodiodes (APDs) and pnCCDs\cite{har08}, and there
is every expectation that it will 
extend towards 2\micron\ over the next few years. 
A high time resolution instrument is essentially an imaging device
with a fast detector, and so  
there would be very little additional opto-mechanical
development required to include such a capability in MICADO's auxiliary
arm.

\end{description}



\end{document}